# Dispersed multi-walled carbon nanotubes in polyvinyl butyral matrix for transparent ionic conductive films

Mykola O. Semenenko[1,2,3], Sergii O. Kravchenko[1], Nadiia V. Siharova[4], Olha V. Pylypova[5], Mariya I. Terets[4], Oleksandr S. Pylypchuk[2], Mariia V. Voitovych[1], Tetiana Yu. Obukhova[3], Taisiia O. Kuzmenko[2,5], and Andrey Sarikov[1,3,5]

[1]V. Lashkaryov Institute of Semiconductor Physics of the National Academy of Sciences of Ukraine, 41 Nauky Avenue, 03028 Kyiv, Ukraine

[2]Institute of Physics of the National Academy of Sciences of Ukraine, 46 Nauky Avenue, 03028 Kyiv, Ukraine

[3]National Technical University of Ukraine "Igor Sikorsky Kyiv Polytechnic Institute", 37 Beresteiskyi Avenue, 03056 Kyiv, Ukraine

[4]Chuiko Institute of Surface Chemistry of the National Academy of Sciences of Ukraine, 17 General Naumov Street, 03164 Kyiv, Ukraine

[5]Educational Scientific Institute of High Technologies, Taras Shevchenko National University of Kyiv, 4-g Hlushkova Avenue, 03022 Kyiv, Ukraine

Corresponding author's email: semandko@gmail.com

**Abstract**

In this work, we develop methods for increasing the dispersion degree of agglomerated multiwalled carbon nanotubes with subsequent introduction of them into polyvinyl butyral to create transparent conductive films. The influence of proton and a-proton solvents in combination with potassium triiodide ($KI_3$) as a redox component for oxidation of the multiwalled carbon nanotubes surface, which reduces agglomeration due to electrostatic repulsion, is investigated. It is demonstrated that a-proton solvent cyclohexanone ensures a smaller size of the agglomerates (30-300 nm, with a maximum of ~145 nm) compared to proton solvent propyl alcohol (100-3000 nm, with a maximum of ~920 nm). The reduced aggregation is associated with the formation of oxygen-containing functional groups (C=O, C-O, C-O-C, and COO), which increase electrostatic stabilization. The impedance analysis showed that the constant component of the conductivity in the samples with multiwalled carbon nanotubes and a-proton solvent shifts to frequencies of ~$10^4$ rad/s after the addition of the redox component, which indicates the formation of ion-conducting channels and stabilization of the jump charge transfer.

**1. Introduction**

Carbon nanotubes (CNTs) find their application in many fields of science and technology due to their exceptional properties such as large surface area, hollow structure, hydrophobicity, ability to chemical modification of the surface, and high electronic conductivity, mechanical strength and thermal conductivity [1, 2].

Fabrication of CNT-based transparent conductive films (TCFs) is one of the most promising applications of CNTs for manufacturing various electronic and optical devices [1, 3, 4]. In general, TCFs include (i) conductive polymers, (ii) thin metal films, and (iii) films containing metal or carbon nanostructures. Among them, conductive polymers are characterized by distinct electrical, optical, and mechanical properties. However, their electrical conductivity is subject to degradation over time due to environmental influences such as humidity, high temperature, and ultraviolet radiation. Thin metal films and the films with metal nanostructures, although exhibiting high electrical conductivity, are easily oxidized. Hence, the films based on carbon nanomaterials, such as CNTs or graphene, are of particular interest due to their good electrical (low resistance), optical (high transparency), mechanical (high strength) characteristics, and high chemical inertness [4].

CNTs have unique physical and chemical properties making them a key component in developing new polymer based composites. By adding CNTs to polymer matrices, it is possible to control the polymer conductivity, strength, flexibility and heat resistance and in this way obtain materials with improved mechanical, thermal, optical, and electrical characteristics. The polymer in such materials acts as a dispersant that facilitates separation of the nanotubes, as well as a matrix to create a composite film. The main challenge is to achieve uniform CNT distribution in the polymer matrix, for which the interfacial adhesion between the CNTs and the polymer matrix plays the major role [5]. To achieve uniform distribution of the CNTs in the polymer matrix during composite manufacturing, the polymer is first added to a proton or a-proton solvent followed by adding CNTs. The mixture of the polymer solution with the dispersed CNTs is used then to prepare conductive transparent films by various coating processes [1, 3, 4].

Solutions with dispersed CNTs are obtained in three main ways: (i) direct dispersion of the CNTs in organic solvents, (ii) dispersion of the CNTs using high-molecular-weight compounds (HMCs) and surfactants (SAs) [6], and (iii) dispersion of the CNTs by introducing new functional groups into the CNT structure [1]. Direct dispersion of the CNTs in an organic solvent is a simple and acceptable method. However, it can be implemented only at low CNT concentrations, which is impractical [4]. Adding HMCs/SAs during CNT dispersion or forming functional groups, such as carboxyl and hydroxyl ones, on the CNT surface helps to increase the degree of the CNT dispersion. Additionally, treatment in an ultrasonic bath may be used [7].

CNTs are difficult to disperse in neat organic solvents, in particular, ethanol. Even after ultrasonic treatment, the CNTs remain highly aggregated. To improve the degree of the CNT dispersion, polyvinyl butyral (PVB) is added to the organic solvents. This substance is widely used in production of safety glasses, adhesives for paints, ceramic films, inks, electronic materials, as an intermediate layer in architecture (facades and construction), in glass lamination, especially for car windshields [8] and in sealing of solar cells [9]. PVB is characterized by a high transparency and elasticity. This material can be stretched up to 2-5 times of its original size before rupture. It has also strong adhesion to surfaces, high resistance to ultraviolet radiation, and retention of properties at high and low temperatures. Adding PVB to CNT solutions made it possible to obtain transparent, flexible and durable films with a moderate electrical conductivity. For example, the conductivity and transmittance (at 550 nm) of the CNT/PVB composite films with 50 and 80 wt% CNT were reported to be 8.81 and 717 S m$^{-1}$, and 82% and 68%, respectively [10].

The goal of this work is to develop methods of increasing the dispersion degree of agglomerated CNTs with their subsequent addition to polyvinyl butyral to create transparent conductive films. The investigations are intended to study the influence of a proton or an a-proton solvent on the dispersion CNT degree. At this, $KI_3$ ions will be added as a redox component for oxidizing the CNT surface, which will increase the electrostatic repulsion between individual CNTs and decrease of the size of agglomerated particles in the PVB matrix.

## 2. Methods

CNTs were prepared by chemical vapor deposition on a metal substrate in a fluidized bed mode with polypropylene as a carbon source [11].

Mixtures of 0.02 g of the CNTs and 0.02 g of PVB (equal mass content of the components, mass %) were dissolved in 10 ml of organic solvents (neat organic solvents) listed in Table 1 and kept in an ultrasonic bath (50 W) for 1 h at room temperature. The solvents used had different values of the hydrophile-lipophile balance (HLB) that defines hydrophilicity (lyophilicity) of the solvent molecules and characterize their affinity for the dispersed particles. HLB is determined by calculating the percentage of the molecular weight of the hydrophilic and lipophilic parts of the molecule (Griffin's

method [12]:

$$HLB = 20 \cdot \frac{M_h}{M}$$

where $M_h$ is the molecular weight of the hydrophilic part of the molecule and $M$ is the molecular weight of the entire molecule, respectively. The HLB range is from 0 to 20, where $HLB = 0$ corresponds to a completely lipophilic (hydrophobic) molecule and $HLB = 20$ completely hydrophilic (lipophobic) molecule, respectively.

Visual inspection showed the highest color intensity for the solutions prepared using the solvents #2 (cyclohexanone) and #4 (n-propyl alcohol) from Table 1. Such coloring evidences stability of the disperse system, in which the dispersed particles do not settle but remain suspended in the solution. Therefore, all the following work was carried out with these solvents.

Table 1

| № | Solvent | $M_h$ | M | HLB | Dipole Moment, D [I1] ([I2]) |
|---|---|---|---|---|---|
| 1 | Cyclohexane | 0 | 84.16 | 0 | 0 |
| 2 | Cyclohexanone | 28.01 | 98.15 | 5.7 | 3.06 |
| 3 | Ethyl Alcohol | 17.01 | 46.07 | 7.38 | 1.66 |
| 4 | n-Propyl Alcohol | 17.01 | 60.09 | 5.66 | 1.68 (3.09) |
| 5 | n-Butyl Alcohol | 17.01 | 74.12 | 4.78 | 1.66 (1.75) |
| 6 | n-Amyl alcohol | 17.01 | 88.15 | 3.86 | 1.66 (1.70) |

[I1] https://www.stenutz.eu/chem/class.php

[I2] https://macro.lsu.edu/Howto/solvents/dipole%20moment.htm

An anionic surfactant (anionic surfactant SDS, sodium dodecyl sulfate, S) or a cationic surfactant (cationic surfactant CTAB, cetyltrimethyl ammonium bromide, C) was added to the initial CNT-PVB solution in a mass ratio of 1% relative to the PVB.

Separately, the CNTs were functionalized by oxidizing their surface. For this purpose, 5 mg of the CNTs and 5 mg of PVBs were added to 15 ml of cyclohexanone and kept in an ultrasonic bath for 20 min. After that, 41.6 mg of KI and 61.6 mg of $I_2$ were added, followed by sonication for 20 min. This solution is the initial one. It is dissolved then in cyclohexanone 1/2 (5 ml of the initial solution is mixed with 5 ml cyclohexanone) and 1/10 (1 ml of the initial solution is mixed with 9 ml of cyclohexanone).

Size distribution of the dispersed particles in the solutions was determined by dynamic light scattering (DLS) using a Malvern Instruments (UK) device equipped with a He-Ne laser ($P_{max} = 4$ mW, $\lambda = 632.8$ nm). The detector position was at an angle of 173°. The volume of the sample solution was fixed at 1 ml. Each measurement was repeated 10 times and the results were averaged.

Transmittance spectra in the ultraviolet and visible (UV-Vis) range of the CNT disperse solutions were measured at the wavelengths of 190 to 1000 nm using a UV-1800 spectrophotometer.

Infrared (IR) transmission and reflection spectra were measured at room temperature using a PerkinElmer Spectrum BX FTIR spectrometer. The measurements were carried out in the spectral range of 350 to 7800 $cm^{-1}$ with the resolution of 4 $cm^{-1}$ and the measurement accuracy of 0.5 %. A clean glass substrate was used as a reference sample.

Capacitance-frequency characteristics were measured in an automatic mode in a parallel equivalent circuit configuration CpD using an R&C LX200 meter. The measurements were performed in the frequency range from 4 Hz to 500 kHz by recording frequency dependences of capacitance and dissipation factor ($D$, equivalent to the dielectric loss tangent, tan δ). The excitation signal amplitude was set to 0.1-1 V to ensure stable and accurate readings across the entire measurement range.

**3. Results and discussions**

Fig. 1 shows the size distributions of the dispersed particles determined by the DLS for the initial solutions as well as for the systems containing surfactants. Dissolution of the CNTs in propyl alcohol provides the size distribution at larger values, namely between 100 and 3000 nm, with a clear peak at about 920 nm (see Fig. 1a). As can be seen from Fig. 1b, the CNT size distribution in the solution in cyclohexanone ranges from 30 to 300 nm with a clear main peak at 145 nm and an additional peak at about 70 nm. An analogous size distribution was observed in [13]. Addition of both anionic and cationic surfactants to the solutions in propyl alcohol shifts the peak of the CNT distribution to about 1480 nm, and small particles almost completely disappear (see Figs. 1b and 1c). Unlike this, no pronounced effect on the size distribution is obtained for the case of the solutions in cyclohexanone (see Figs. 2b and 2c). Addition of equivalent amounts of KI and $I_2$, which leads to formation of potassium triiodide $KI_3$, to the CNT-PVB causes oxidation of the CNT surface and appearance of functional oxygen-containing groups. At this, the size distribution of the dispersed particles shifts toward smaller values with the maximum at ~10 nm, see Fig. 2.

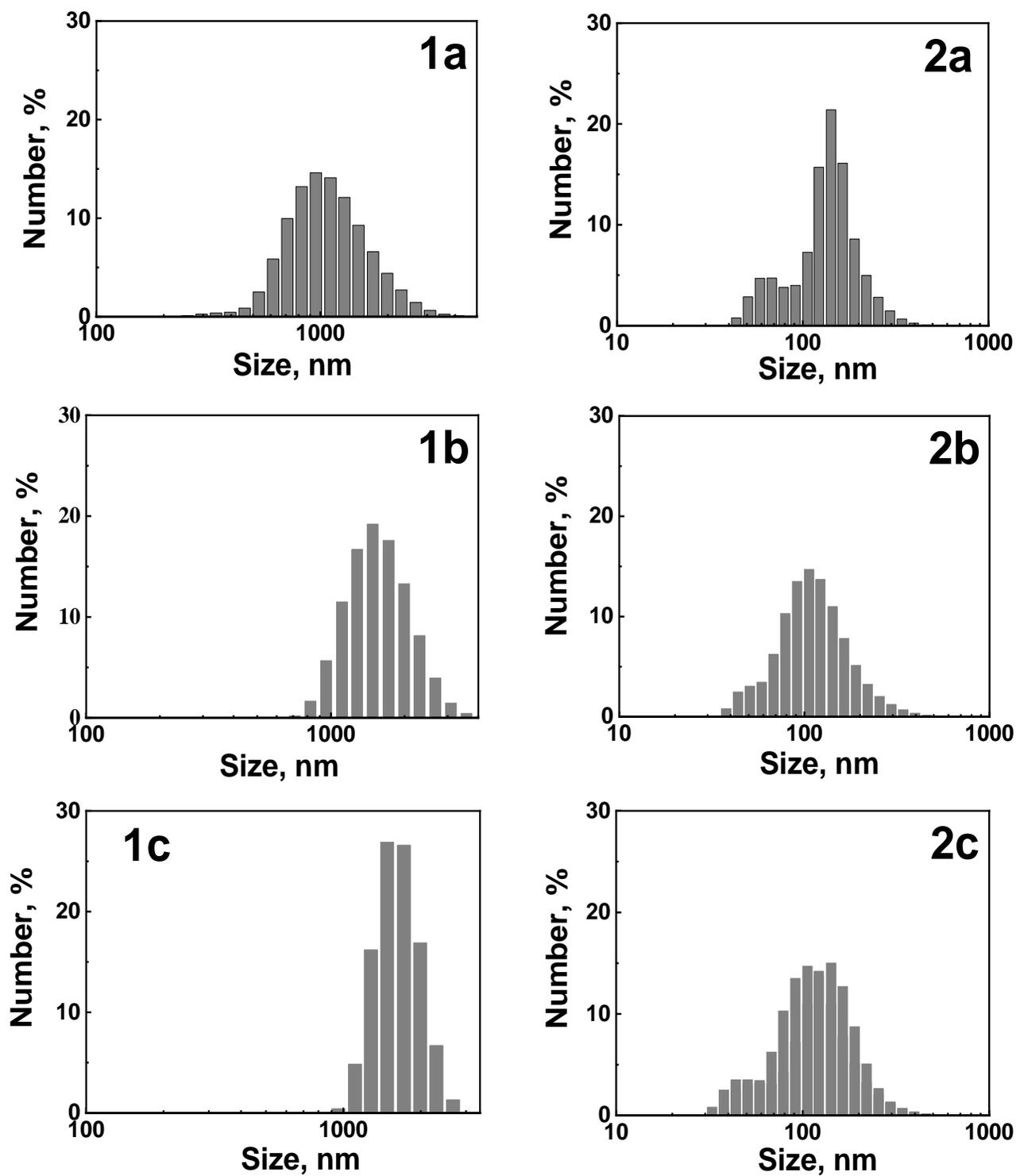

Fig. 1. Size distributions of CNT particles in disperse solutions with propyl alcohol (1) and cyclohexanone (2): (a) – initial CNT–PVB solution; (b) – after adding surfactant SDS; (c) – after adding CTAB cationic surfactant.

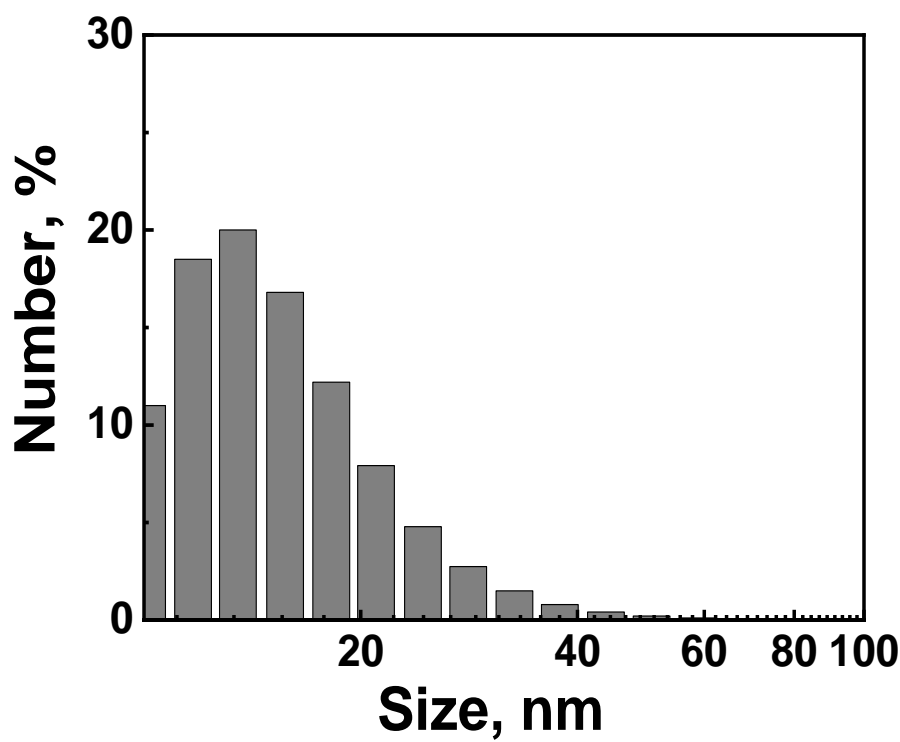

Fig 2. Size distribution of particles in a disperse CNT solution after addition of cyclohexanone and $KI_3$.

One may assume that addition of $KI_3$ induces dispersion of the agglomerated CNTs to isolated CNTs due to surface oxidation. To confirm this, one should consider the factors influencing the dispersion degree of particles in a disperse system. In general, four factors influence the stability of the disperse system, in particular with CNTs: (i) electrostatic factor due to the zeta potential on the particle surfaces, (ii) adsorption-solvation factor caused by action of the surfactants, (iii) structural and mechanical factor, and (iv) entropic factor caused by action of the high-molecular-weight compounds. In our case, PVB provides the structural-mechanical and entropic factors of the stability of the disperse system. Similar to surfactants, PVB is a dispersing agent (dispersant), i.e. a compound added to a disperse solution to improve separation of dispersed solid particles in a liquid medium in order to prevent their aggregation. A PVB molecule has affinity for both CNT solid particles (dispersed phase) and molecules of the solvent containing these particles (disperse medium).

A PVB molecule contains hydrophobic (butyl (Bu) and acetyl (Ac)) and hydrophilic (hydroxyl (OH)) groups (see Fig. 3).The quantitative ratio between the Bu, Ac, and OH groups in the PVB influences its physical and chemical properties, such as solubility, miscibility, viscosity, strength, flexibility, and ability to interact with other compounds [14]. As can be seen from our results, the solutions with the most dispersed particles were obtained with propyl alcohol and cyclohexanone.

These solvents have similar HLB values. It can be assumed therefore that the PVB surface should have a comparable HLB value.

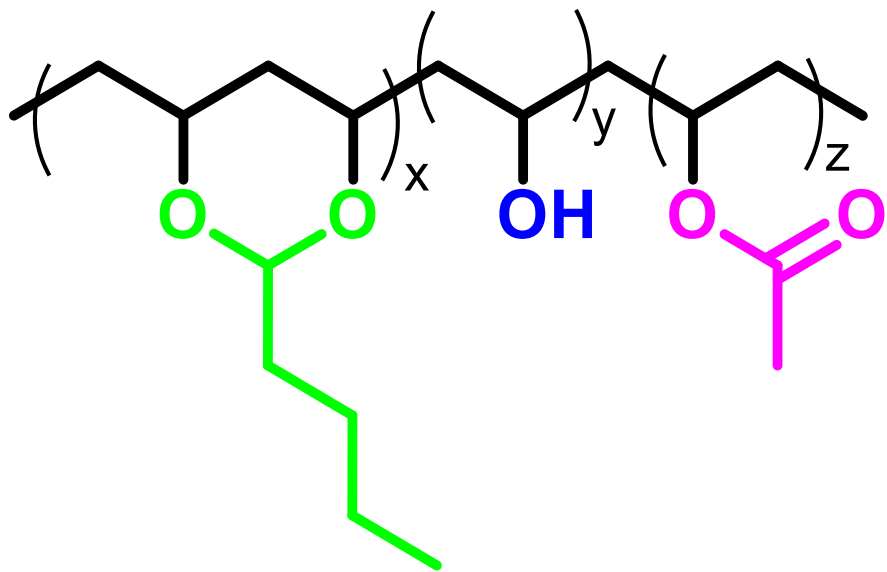

Fig. 3. Structure of a PVB molecule. Coloring: green – Bu, violet – Ac, blue – OH.

As mentioned above, aggregation of CNTs in a dispersed phase takes place due to the van der Vaals forces between $sp^2$-hybridized carbon atoms of closely spaced CNTs. On the other hand, dispersion of the CNTs is enabled by electrostatic repulsion forces due to the formation of oxygen-containing functional groups on the CNT surface upon interaction of carbon atoms with oxygen atoms (oxygenation), in particular, the $sp^2$-hybridized oxo (C=O) group, and $sp^3$-hybridized oxy (C-O), epoxy (C-O-C), and carboxy (COO) groups. Hence, the oxygen-containing functional groups formed after adding potassium triiodide ($KI_3$) to the disperse systems with CNTs and PVB as a dispersant are likely to contribute to the electrostatic repulsion. This is manifested in the reduction of the size distribution in the disperse CNT solution after treatment with $KI_3$ (see Fig. 2).

Characteristic UV–Vis spectra of the CNTs are shown in Fig. 4. As can be seen from this figure, these absorption spectra exhibit characteristic peaks at 305 nm and 916 nm. The peak at 305 nm corresponds to absorption of π-plasmons associated with collective excitations of π-electrons at wavelengths of 310-155 nm (4.0-8.0 eV) [15]. Additional characteristic peaks at 430 nm and 540 nm are observed for the propyl alcohol based solutions, and an additional peak at 384 nm is observed for the solutions prepared with cyclohexanone. The typical absorption peak at 916 nm reveal the characteristics of electronic transitions corresponding to a metallic semiconducting carbon nanotube sample [16].

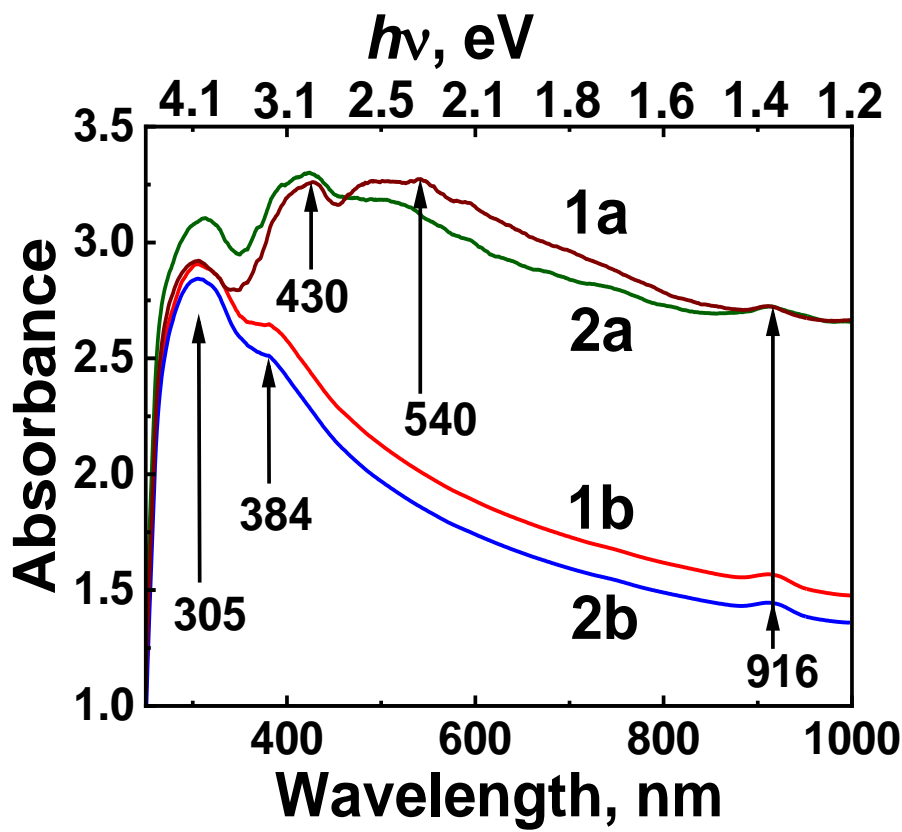

Fig. 4 UV–Vis spectra. 1a – CNT–PVB propyl alcohol SDS; 2a – CNT–PVB propyl alcohol CTAB. 1b – CNT–PVB cyclohexanone SDS; 2b – CNT–PVB cyclohexanone CTAB.

We calculated the optical bandgap energy $E_g$ of the CNTs from the UV-Vis spectrum using the Tauc equation, which provides a relationship between the absorption coefficient [17] and the incident photon energy:

$$(\alpha h\nu)^{1/n} = A(h\nu - E_g)$$

Here, α is the absorption coefficient, $n$ is a constant dependent on the type of electron transitions ($n = 1/2$ for direct allowed transitions), $h$ is the Planck's constant, and $v$ is the frequency of light, respectively.

To calculate the $E_g$ value for the CNTs, we built a plot $(\alpha h\nu)^2$ vs. incident photon energy ($h\nu$),

which is presented in Fig. 5. The $E_g$ value was estimated by extrapolating the straight-line segment of the plot $(\alpha h\nu)^2$ vs $(h\nu)$ to a zero value of the absorption coefficient for direct allowed optical transitions. From Fig. 5a, the $E_g$ for the CNTs dissolved in cyclohexanone with SDS and CTAB was found to be 2.77 eV and 2.86 eV, respectively. From Fig. 5b, the $E_g$ value of the CNTs dissolved in propyl alcohol with SDS and CTAB was found to be 2.54 eV, in agreement with the respective experimental result [17].

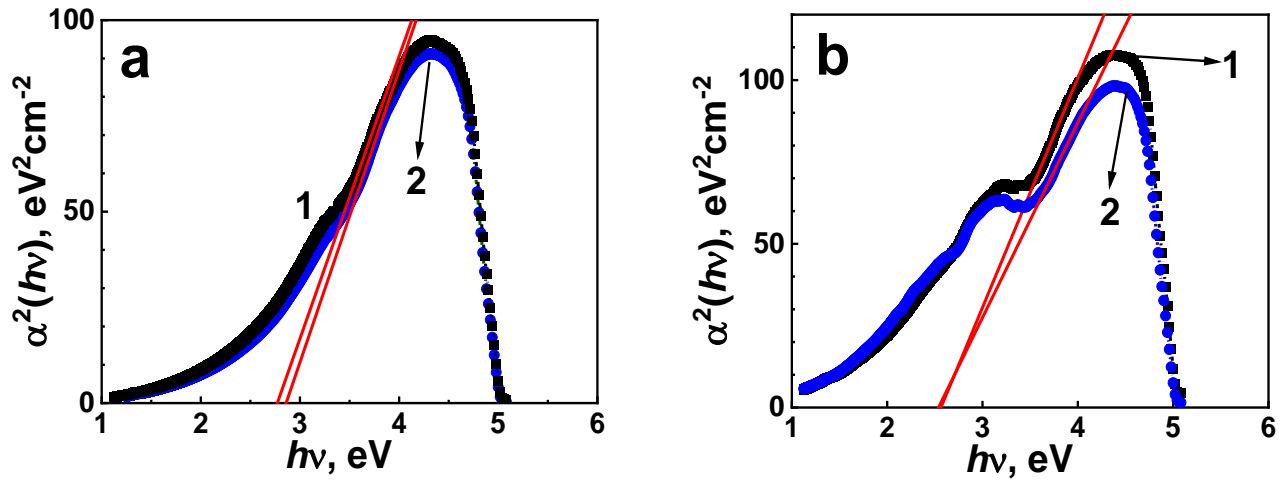

Fig. 5. Tauc's plot for the absorption coefficient of CNT-PVB in different solvents (1 – SDS; 2 – CTAB): a) propyl alcohol; b) cyclohexanone.

Fig. 6 shows IR absorption spectra of the films of CNTs in propyl alcohol and cyclohexanone in the range of 2800-3100 cm$^{-1}$. These spectra contain the bands with the maxima at ~2850 cm$^{-1}$, ~2870 cm$^{-1}$, ~2920 cm$^{-1}$, and ~2960 cm$^{-1}$. As can be seen from this figure, the positions of the peaks and the shape of the IR absorption spectra of the films of CNTs in propyl alcohol and cyclohexanone are substantially different. For the CNTs in the propyl alcohol, the peaks near ~2920 and ~2960 cm$^{-1}$ are pronounced, while for the CNTs in cyclohexanone, they are shifted in the low-frequency region to ~2850 and ~2920 cm$^{-1}$, respectively.

Using the maximum positions of the recorded IR bands, we can identify the presence and the type of the interatomic bonds formed in the samples [18]. In the frequency range of 2800-3100 cm$^{-1}$, the known IR absorption peaks are associated with CH$_n$ bonds that can be either sp$^3$- (2850-2960 cm$^{-1}$, aliphatic hydrogen) or sp$^2$-hybridized (2960-3100 cm$^{-1}$, aromatic hydrogen) [19-22]. For both the films with the CNTs in propyl alcohol and cyclohexanone, the IR spectra show well-defined bands corresponding to vibrations of the aliphatic sp$^3$-hybridized CH$_n$ bonds.

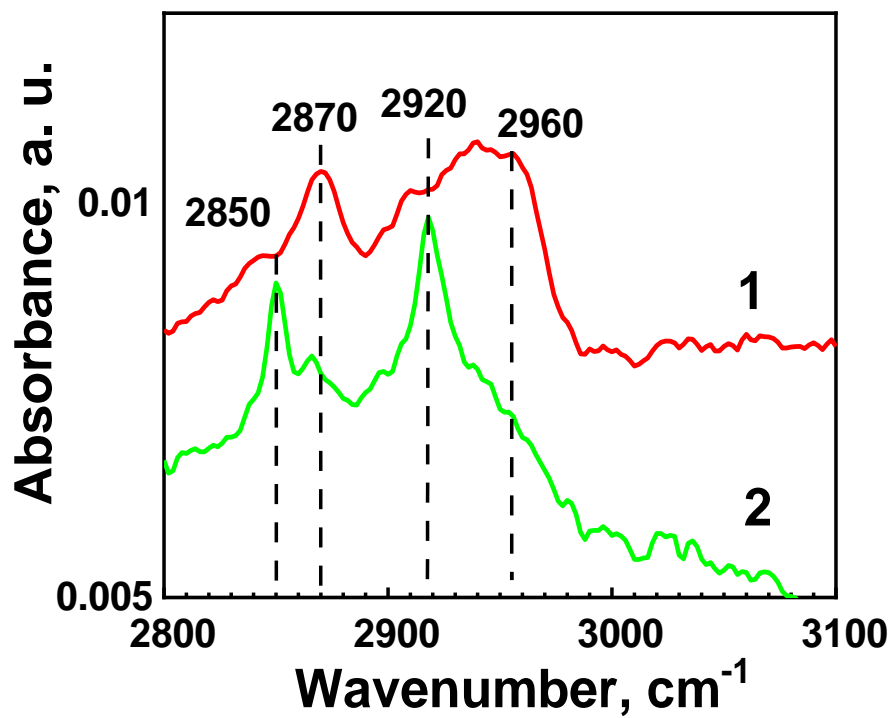

Fig. 6. IR absorption spectra of the CNT films formed with propyl alcohol (1) and cyclohexanone (2).

For a more detailed and in-depth analysis of the structure of the studied films, we performed a mathematical treatment of the shape of the IR absorption band at C–H vibrations with identification of the elementary components (peak positions ν and half-width w) as described in [20, 22]. This allowed us to more accurately determine the structural state of carbon and hydrogen in the films and in this way obtain additional information about the characteristic $CH_n$ (0 < n < 4) chemical bonds and the contribution of symmetric and asymmetric vibrations in the films. Decomposition of the IR absorption spectra of the films with CNTs in propyl alcohol and cyclohexanone in the frequency region of aliphatic C–H vibrations allowed us to identify Gaussian components with the peaks at ~2830 $cm^{-1}$, ~2850 $cm^{-1}$, ~2872 $cm^{-1}$, ~2900 $cm^{-1}$, ~2925 $cm^{-1}$, and ~2955 $cm^{-1}$, as shown in Fig. 7. All the obtained parameters (peak positions ν and half-width sw), their interpretation, and contribution of each component (band area, %) to the integrated IR spectrum are presented in Table 2.

The elementary bands, the maximum positions of which we have identified, are inherent in $CH_2$, $CH_3$ and CH bonds in the $sp^3$-hybridized state. The bands 1 and 2 correspond to the symmetric vibrations of hydrogen and carbon atoms in the $CH_2$ and $CH_3$ groups, and the bands 4 and 5 are associated with their asymmetric vibrations, respectively. The band 3 is caused by CH aliphatic vibrations, which is confirmed in [21, 23, 24]. The spectrum of the film with the CNTs in propyl alcohol showed an additional peak at ~2830 cm$^{-1}$ (band 0 in Fig. 7a), which is absent in the absorption spectrum of the CNTs in cyclohexanone. The corresponding band may be attributed to symmetrical vibrations in methylene groups $CH_2$ with $sp^3$-hybridizaton [21, 24, 25].

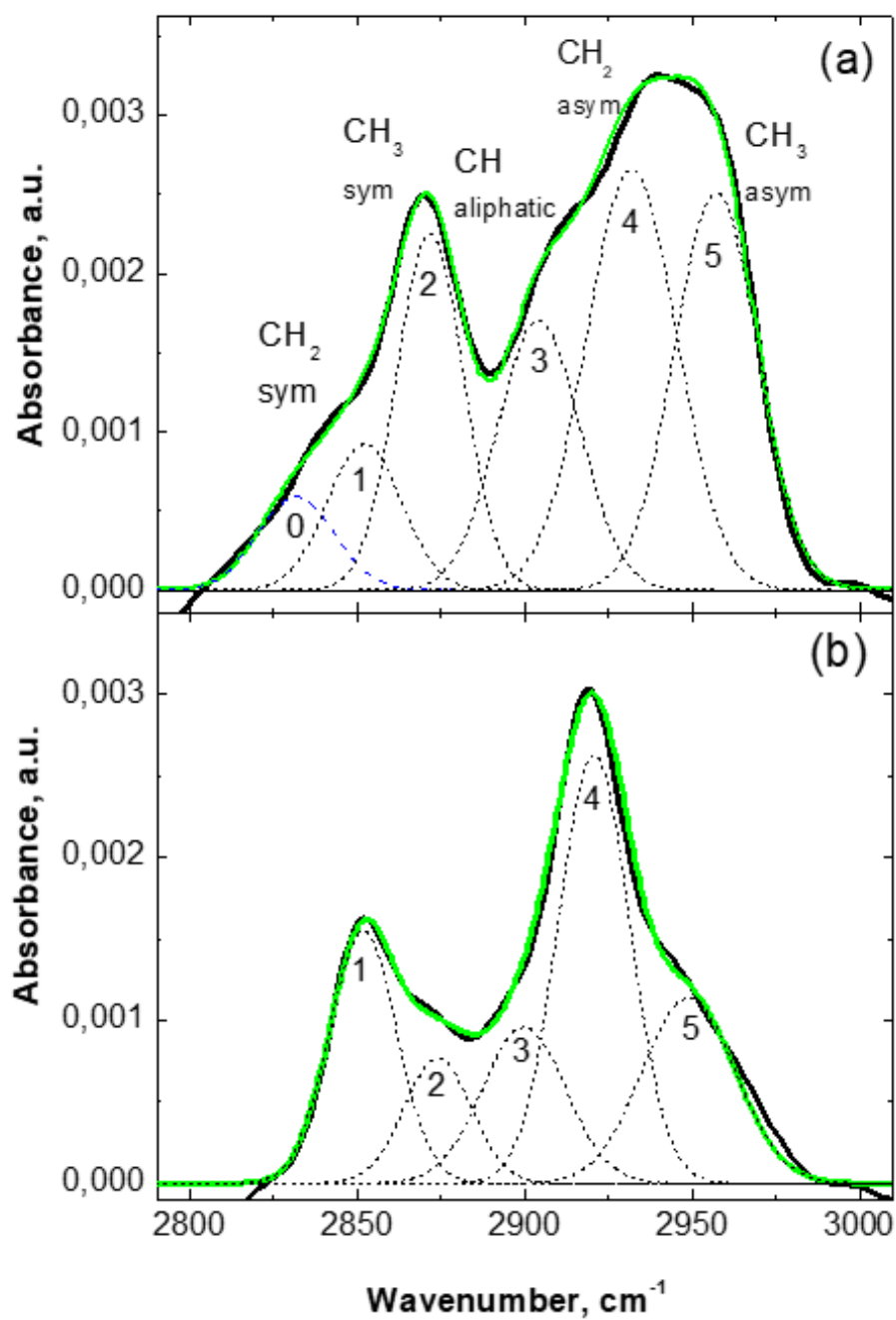

Fig. 7. IR absorption spectra of the CNT films formed with propyl alcohol (a) and cyclohexanone (b) deconvoluted into Gaussian components corresponding to C–H bond vibrations.

Table 2. Parameters of elementary absorbance bands of the propyl alcohol (1) and cyclohexanone (2)

| Absorbance bands | Aliphatic C-H stretching (2800–3000 cm$^{-1}$), % | | | | | |
|---|---|---|---|---|---|---|
| | 0 | 1 | 2 | 3 | 4 | 5 |
| | CH$_2$ symmetric | CH$_2$ symmetric | CH$_3$ symmetric | CH aliphatic | CH$_2$ asymmetric | CH$_3$ asymmetric |
| Position ν (cm$^{-1}$) | 2830±2 | 2850±4 | 2872±2 | 2900±2 | 2925±5 | 2955±5 |
| Bandwidth $w$ (cm$^{-1}$) | 21±1 | 20±2 | 20±2 | 25±2 | 25±3 | 25±3 |
| **propyl alcohol (1)** | 5.3 | 8 | 17.6 | 16.4 | 28.1 | 24.6 |
| **cyclohexanone (2)** | | 20 | 9.7 | 15 | 35.7 | 19.4 |

Taking into account that the area of the absorption band at the respective vibration frequency is proportional to the amount of chemical bonds [18], a comparative analysis of the number of the C–H bonds in the CNT films formed with propyl alcohol and cyclohexanone can be performed. Calculation of the contributions of the absorbance components (see Table 2) demonstrate that the chemical compositions of the CNTs films formed with propyl alcohol and cyclohexanone are somewhat different. The table shows that CH$_3$ bonds (sp$^3$-hybridization) dominate in the CNT films formed with propyl alcohol (17.6 %), while in the CNT films formed with cyclohexanone they are 1.8 times smaller (9.7 %). At this, more CH$_2$ bonds (sp$^3$-hybridization) form in the CNT films with cyclohexanone. The methylene groups were present in both cyclohexanone and propyl alcohol case, while the methyl groups were present only in the case of propyl alcohol.

We carry out a comprehensive analysis of the electrical impedance of the PVB films dissolved in cyclohexanone and deposited on substrates with added KI+I$_2$, and with added CNTs, followed by film drying. The substrates had two concentric ring-shaped gold contacts used to measure the impedance response. After eliminating parasitic signals, the Nyquist diagrams (see Fig. 8) and the frequency dependences of the conductivity (see Fig. 9) defined in absolute units as $G = \cos(\varphi)/|Z|$, were built. As can be seen from Fig. 8, the Nyquist diagram for the initial film without potassium triiodide ions shows a line inclined to the $Z_{real}$ axis in the low-frequency range, which is characteristic of diffusion-limited charge transfer or Warburg impedance behavior. On the other hand, a flattened semicircle at higher frequencies points to interfacial polarization, in particular, according to the Maxwell-Wagner mechanism. In the samples obtained from the solutions with lower PVB concentration, this behavior is somewhat different. The Nyquist diagram shows a semi-arc from zero to a quarter of a circle. This indicates the presence of a capacitive response with the time constant similar to all the samples, in accordance with the Maxwell-Wagner model. That is, interface polarization mechanism, e.g. at the PVB matrix/ions interface acts here. It should be noted that the dependences of the imaginary dielectric function on the circular frequency do not show any relaxation peaks (not presented here, since they are informative only when relaxation peaks are available and can be used for estimating the relaxation time of the relaxer). Therefore, the interfacial polarization according to the Maxwell-Wagner mechanism is dominant. Addition of $KI_3$ ions does not change the semicircle in the Nyquist diagram. Moreover, there is no peak in the frequency dependence of the imaginary part of the dielectric function or a Z-shaped frequency dependence of the real part of the dielectric function, which points to absence of relaxation processes.

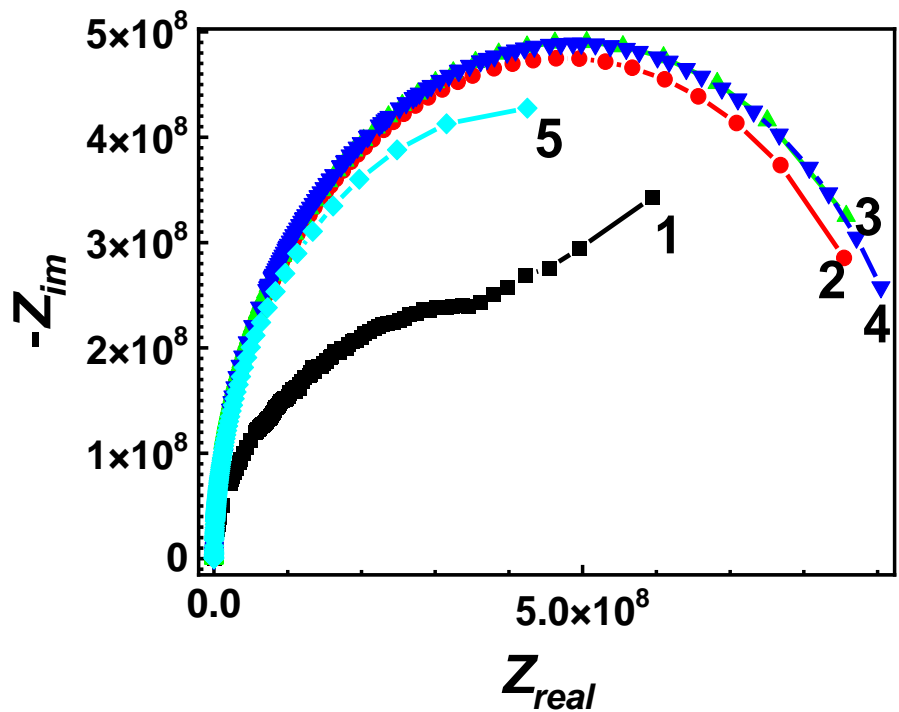

Fig. 8. Nyquist diagram of the PVB films dissolved in cyclohexanone and deposited onto substrates in different concentrations: (1) – initial; (2) – 1/2; (3) – 1/10; (4) – with addition of $KI_3$ ions; (5) – with addition of CNT.

We present the frequency dependence of the conductivity $G = \cos(\varphi)/|Z|$ in double logarithmic coordinates (Jonher projection) to unveil possible changes in the conduction mechanisms (see Fig. 9). As can be seen from Fig. 9, all the curves show a slow increase at low frequencies, which corresponds to the dominance of bulk conductivity with ion migration. At high frequencies, the increase of the curve slope is observed. In general, the behavior of the film conductivity agrees well with the Maxwell-Wagner mechanism. The conductivity is mainly formed by ionic transport, which is enhanced by addition of $KI_3$ ions. The horizontal plateau at low frequencies indicates a steady-state (DC) ionic conductivity, i.e., ions move in the bulk causing charge transfer. At the same time, the Nyquist diagram with a quarter-circle starting from zero (high frequencies) shows the dominance of the capacitive response. This response may be either due to interfacial polarization in the near-electrode region or due to charge accumulation at grain boundaries. This means that although the conductivity is ionic in nature, a significant portion of the charge carriers accumulate and do not immediately pass through the structure. These carriers have a certain delay time manifested as the capacitance. The Nyquist diagram does not change its shape with increasing frequency, i.e., the polarization mechanism remains acting. However, the conductivity plot begins to go upward, and the frequency-dependent conductivity (AC) emerges, which may evidence a hopping type of charge transfer (jumping of ions between the localized states).

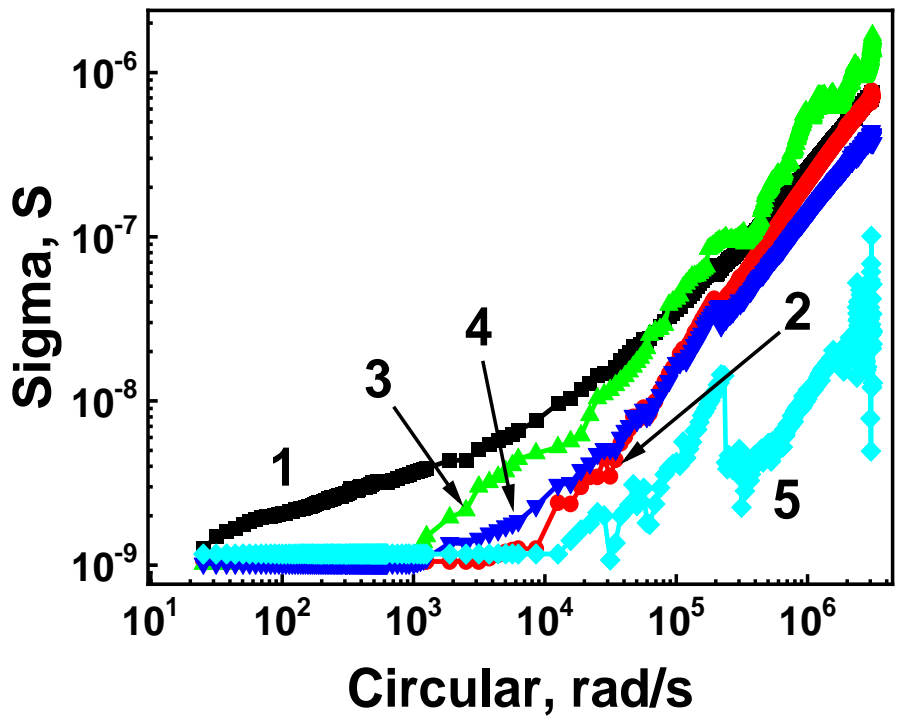

Рис. 9. Jonher projections for dependences of the film absolute conductivity on the circular frequency for the PVB films dissolved in cyclohexanone and deposited onto substrates in different concentrations: (1) – initial; (2) – 1/2; (3) – 1/10; (4) – with addition of $KI_3$ ions; (5) – with addition of CNT.

Addition of the CNTs to the PVB film formed with $KI_3$ ions causes a significant decrease in the impedance and an increase in the conductivity, which is shown in the Nyquist diagram by the smallest semicircle (see Fig. 8). At the same time, the frequency dependence of the conductivity shows a sawtooth behavior in the high-frequency region, which points to the intermittent nature of charge transfer due to the formation of a clustered CNT conductive network. The DC component of the conductivity increases and remains constant up to the frequencies of ~$10^4$ rad/s. This makes a contrast to the samples without the CNTs, for which the DC component of the conductivity disappears at the frequencies of about $10^3$ rad/s, indicating a more efficient ionic conductivity. This is explained by the fact that ions have more time to relax and pass through the interfacial regions. The CNTs create topological paths facilitating ion movement (through defects, intertube gaps, surface interaction with the matrix, etc.) The high value of the phase (>45°) in the body of the Bode phase diagram (we do not present it here since the frequency dependences of the phase are monotonous) confirms the prevalence of the capacitive response over the ionic hopping mechanism at high frequencies for all the samples. At this, the best ionic conductivity at low frequencies is characteristic of the disperse systems with CNTs. Here, the delayed DC component (horizontal straight line, see Fig. 9) points to the existence of effective ion-conducting channels.

Therefore, we may draw the following conclusions based on the analysis of the conductivity, impedance, and phase plots. In the Nyquist diagram (see Fig. 8), only the initial sample (init) exhibits a linear segment in the low-frequency region, which corresponds to diffusion processes and the Warburg resistance. In all the other samples, including those with the CNTs, semicircles are observed, which are characteristic of conductivity through localized states with transient impedance formed due to a limited number of ion channels or presence of barriers between the phases. Absence of the straight segments in the Nyquist diagram for the CNT samples shows that addition of CNTs does not cause transition from the ionic to the electronic conductivity, but instead leads to a formation of localized conductive paths.

**4. Conclusions**

In this work, we have developed methods for increasing the dispersion degree of agglomerated multiwalled carbon nanotubes with subsequent introduction of them into polyvinyl butyral to create transparent conductive films. The obtained results show that use of cyclohexanone as a hydrophobic solvent ensures a more efficient dispersion of multiwalled carbon nanotubes in a solution as compared to propyl alcohol, which is hydrophilic. The size of the nanotube aggregates in the cyclohexanone was 30-300 nm with a distinct peak at 145 nm. In contrast, the particle size in the propyl alcohol was much larger (100-3000 nm with a peak at 920 nm), implying nanotube aggregation. This can be explained by the difference in the hydrophilic-hydrophobic balance of the solvents. Cyclohexanone, being a more hydrophobic solvent, interacts more effectively with the nanotubes, stabilizing them in the solution. Since propyl alcohol is a-proton solvent, i.e., capable of forming hydrogen bonds, the degree of CNT agglomeration will be higher. At the same time, cyclohexanone is an a-proton solvent, i.e., it lacks the ability to form hydrogen bonds. Therefore, use of the latter results in a dispersed system with smaller agglomerated CNT particles.

The IR absorption spectra of the investigated disperse systems show clearly resolved bands in the range of 2800-2950 cm$^{-1}$ corresponding to the sp$^2$- and sp$^3$-bonds in cyclohexanone. Presence of these bonds indicate preservation of the nanotube structure and their stability in the solution. $CH_3$ bonds (sp$^3$-hybridization) dominate in the CNT films formed with propyl alcohol (17.6 %), while in the CNT films formed with cyclohexanone they are 1.8 times smaller (9.7 %). At this, more $CH_2$ bonds (sp$^3$-hybridization) form in the CNT films with cyclohexanone. The methylene groups were present in both cyclohexanone and propyl alcohol case, while the methyl groups were present only in the case of propyl alcohol. Aggregation of multiwall carbon nanotubes in a dispersed phase takes place due to the van der Vaals forces between sp$^2$-hybridized carbon atoms of closely spaced multiwall carbon nanotubes. On the other hand, dispersion of the multiwall carbon nanotubes is enabled by electrostatic repulsion forces due to the formation of oxygen-containing functional groups on the multiwall carbon nanotubes surface upon interaction of carbon atoms with oxygen atoms sp$^2$-hybridized oxo (C=O) group, and sp$^3$-hybridized oxy (C-O), epoxy (C-O-C), and carboxy (COO) groups. Hence, the oxygen-containing functional groups formed after adding KI3 to the disperse systems with multiwall carbon nanotubes and polyvinyl butyral as a dispersant are likely to contribute to the electrostatic repulsion. This is manifested in the reduction of the size distribution in the disperse multiwall carbon nanotubes solution after treatment with KI$_3$.

Analysis of the frequency dependence of the impedance shows that the DC component of the conductivity in the samples with carbon nanotubes is delayed to the frequencies of ~10$^4$ rad/s, evidencing effective formation of ion-conducting channels and stabilization of the hopping charge transfer. Introduction of nanotubes does not provide a through electronic conductivity, but facilitates localized charge transfer in the heterogeneous composite medium through effective ion-conducting channels.

**Author contributions**

**M.O.S.:** conceptualization, formulation of research tasks, analysis of the results; **S.O.K.:** development of technological routes for the formation of solvents and solutions, discussion of results;

**N.V.S.:** literature review, fabrication and characterization of nanotubes, discussion of the results; **O.V.P.:** UV-VIS-IR transmission measurements, analyzing dynamic light scattering (DLS) spectra; **M.I.T.:** fabrication of agglomerated nanotubes; **O.S.P.:** obtaining and analyzing capacitance-frequency characteristics; **M.V.V.:** acquisition and analyzing FTIR spectroscopy results; **T.Yu.O.:** preparation of samples for impedance and conductivity measurements; **T.O.K.:** preparation of samples and measurement of dynamic light scattering (DLS) spectra; and **A.S.:** coordinating the research, discussing the results, and preparing the publication for publication.

**Acknowledgement**

M.O.S., M.V.V. and A.S. acknowledge support of their research by the project 4Ф-2024 "Multilayer structures with organic polymer semiconductor heterojunctions and Si photon crystal based backside reflectors for photoelectric solar energy converters" of the Program of Joint Research Projects of Scientists of the Taras Shevchenko National University of Kyiv and the National Academy of Sciences of Ukraine in 2024-2025. O.S.P., T.O.K. and M.O.S. acknowledge support of their research by the Project No. 5.8/25-П "Energy-saving and environmentally friendly nanoscale ferroics for the development of sensorics, nanoelectronics and spintronics" of the Target Program of the National Academy of Sciences of Ukraine. T.Yu.O., A.S. and M.O.S. acknowledge support by the project "Development of photovoltaic cells and thin-film photoconverters based on organic solar light absorbers" of the Ministry of Education and Science of Ukraine under the state registration No. 0125U001660.